\input{aipcheck}

\documentclass[
    ,final            
  ]
  {aipproc}

\layoutstyle{6x9}

\begin{document}

\title{The Reverse Shock of SNR 1987A}


\author{Kevin Heng}{
  address={JILA, University of Colorado, Boulder, CO 80301-0440}
}

\begin{abstract}
The reverse shock of supernova remnant (SNR) 1987A emits in H$\alpha$ and Ly$\alpha$, and comes in two flavors: surface and interior.  The former is due to direct, impact excitation of hydrogen atoms crossing the shock, while the latter is the result of charge transfer reactions between these atoms and slower, post-shock ions.  Interior and surface emission are analogous to the broad- and narrow-line components observed in Balmer-dominated SNRs.  I summarize a formalism to derive line intensities and ratios in these SNRs, as well as a study of the transition zone in supernova shocks; I include an appendix where I derive in detail the ratio of broad to narrow H$\alpha$ emission.  Further study of the reverse shock emission from SNR 1987A will allow us to predict when it will vanish and further investigate the origins of the interior emission.
\end{abstract}

\maketitle


\section{Introduction: SNR 1987A}

For the past 20 years, supernova remnant (SNR) 1987A has provided a wonderful opportunity to study emission mechanisms, radiative transfer and a myriad of physics for conditions unattainable on Earth.  One such sub-field is the study of high Mach number, collisionless shocks.  The impact of the supernova (SN) blast wave upon ambient medium sets up a double shock structure consisting of a forward and a reverse shock.  In SNR 1987A, the ejecta comprising mostly neutral hydrogen (which exists due to adiabatic expansion cooling) crosses the reverse shock at $\sim$ 12,000 km s$^{-1}$; the excitation and subsequent radiative decay of the atoms result in H$\alpha$ and Ly$\alpha$ emission, readily measured by instruments such as the Space Telescope Imaging Spectrograph (STIS) onboard the {\it Hubble Space Telescope} (see \cite{Heng:2006} and references therein).  

\begin{figure}
  \includegraphics[width=5.5in]{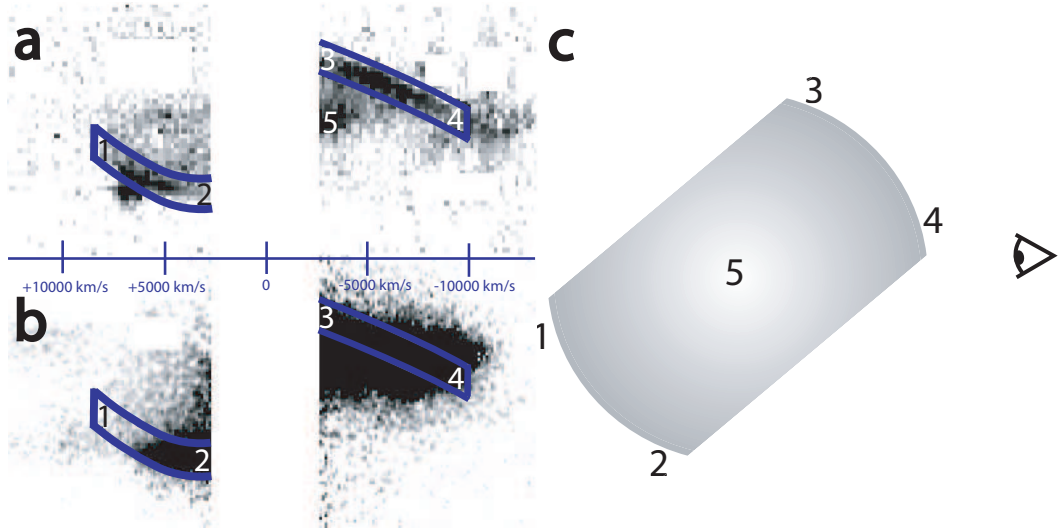}
  \caption{STIS data of reverse shock emission from SNR 1987A and accompanying schematic, taken from \cite{Heng:2006}.  (a) H$\alpha$ surface emission from the reverse shock isolated by masks.  (b) Ly$\alpha$ surface emission with the same masks applied.  (c) Schematic representation of the supernova debris with the boundary being defined by the reverse shock.  For freely-expanding debris, there is a unique correspondence between velocity and the origin of the emission along the line of sight.}
\label{fig:stis}
\end{figure}

In the most recent study of the reverse shock \cite{Heng:2006}, it was found that both H$\alpha$ and Ly$\alpha$ emission exist in two flavors: surface and interior.  In a young, pre-Sedov-Taylor remnant such as SNR 1987A, the freely-streaming debris has a unique velocity for a given radial distance from the SN core, exactly analogous to Hubble flow in an expanding universe.  The projected velocity of the atoms crossing the reverse shock is proportional to the line-of-sight depth of the shock surface from the supernova mid-plane.  It follows that upon impact excitation, the wavelength of the emitted photon is uniquely related to this depth, and the emission streaks in Figure [\ref{fig:stis}] trace out the surface of the reverse shock, thereby warranting the term ``surface emission''.  If one believes this interpretation, then it is apparent from Figure [\ref{fig:stis}] that there is both H$\alpha$ and Ly$\alpha$ emission emerging from beneath the surface of the reverse shock, since at any given frequency or wavelength, flux appears at radial distances smaller than the radius of the shock.  On this basis, we coin the term ``interior emission''.

The shock velocity of SNR 1987A is $\sim 8000$ km s$^{-1}$, since it is the velocity of the atoms in the rest frame of the reverse shock, moving at $\sim 4000$ km s$^{-1}$.  Strong shock jump conditions dictate that the ions are then at a velocity of $\sim 6000$ km s$^{-1}$ in the observer's frame.  Thus, the fast atoms are being converted into slow ions at the reverse shock.  In addition to impact excitation, atoms may also donate their electrons to ions in the shocked plasma (i.e., charge transfer), thereby producing a population of slow atoms.  The subsequent excitation (or charge transfer to excited states) of these atoms results in lower velocity H$\alpha$ and Ly$\alpha$ emission, creating the illusion that these photons originate from beneath the reverse shock surface --- ``interior'' emission.  Both interior and surface emission originate from the same location, but the spectral-spatial mapping is no longer unique.

\section{Balmer-Dominated Supernova Remnants}

Dick McCray and I puzzled over the origins of the interior emission --- he came up with the charge transfer idea, while I sat down and worked out the mathematical details.  Deep into creating a formalism to compute the line intensities and ratios, I stumbled upon an old problem, namely the study of Balmer-dominated SNRs (\cite{CR78}, \cite{CKR80} and \cite{HengMcCray:2007}).  (I call the problem ``old'' because it was posed in the same year I was born.)  These objects are typically much older than SNR 1987A, and are observationally characterized by two-component, Balmer line profiles consisting of a narrow ($\sim 10$ km s$^{-1}$) and a broad ($\sim 1000$ km s$^{-1}$) line.  The former comes from the direct, impact excitation of stationary hydrogen atoms by the SN blast wave, while the latter is a result of charge transfer reactions of these atoms with post-shock ions.  

\begin{figure}
  \includegraphics[width=5.5in]{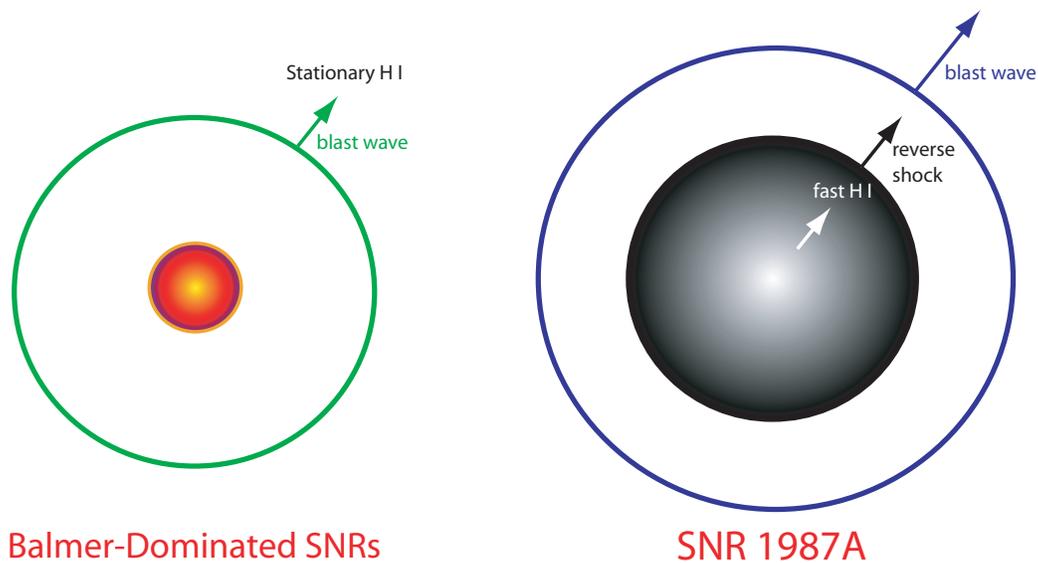}
  \caption{Contrasting the physical situations in ``normal" Balmer-dominated SNRs and SNR 1987A, taken from \cite{Heng:thesis}.}
\label{fig:snrs}
\end{figure}

The terms ``fast'' and ``slow'' are solely a matter of one's frame of reference.  In the frame of the observer, the situation of fast atoms and slow ions in SNR 1987A now gets switched to slow atoms and fast ions in these Balmer-dominated SNRs (Figure [\ref{fig:snrs}]).  The interior and surface emission of the former are the broad and narrow components of the latter.  Nevertheless, the physics of the problem remain the same.  I suddenly realized that I now had the mathematical machinery not only to model the emission lines in SNR 1987A, but to treat this broader class of objects as well.  We generalized the methods of \cite{CKR80} --- we asked the question: can one exhaustively track the fate of a hydrogen atom as it engages in charge transfer and excitation, eventually culminating in impact ionization?  

It turns out that we can if we make certain fairly accurate approximations, allowing us to find simple, analytical formulae for the rate coefficients of these reactions, weighted by how many times the atom undergoes charge transfers; each such event changes the nature of the atomic velocity distribution \cite{HengMcCray:2007}.  By knowing how to compute these rate coefficients, we can in turn compute the probability for each reaction occurring, thereby obtaining the {\it composite} velocity distribution.  These distributions are intermediate between a beam and a Maxwellian, and we thus named them ``skewed Maxwellians''.  We call atoms in such a skewed Maxwellian ``broad neutrals''.  The full width at half-maximum (FWHM) of these velocity distributions is then uniquely related to the shock velocity, provided one knows the temperatures of the electrons and ions.  I have included an appendix describing in detail the derivation of the broad and narrow H$\alpha$ rate coefficients (\S\ref{sect:append}), since the ratio of broad to narrow H$\alpha$ emission is extensively studied in Balmer-dominated SNRs.

\begin{figure}
  \includegraphics[width=5.5in]{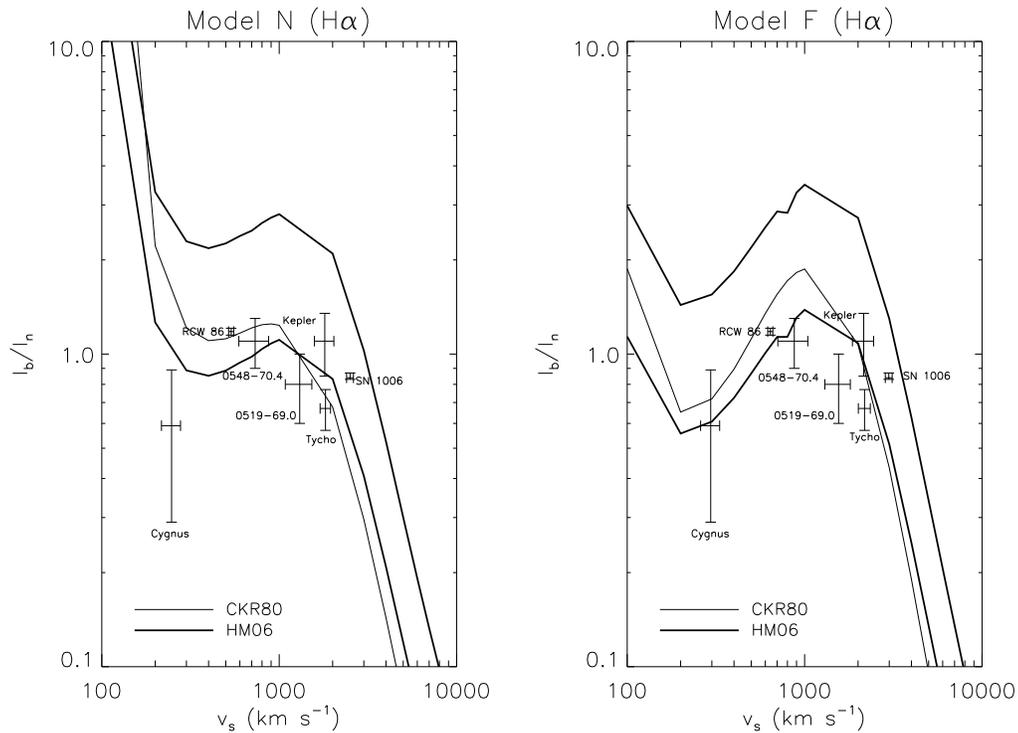}
  \caption{Ratio of the broad to narrow H$\alpha$ emission, $I_b/I_n$, versus shock velocity, $v_s$.  The theoretical predictions by \cite{CKR80} (denoted ``CKR80'') and \cite{HengMcCray:2007} (denoted ``HM06'') are plotted against several data points from various SNRs.  Models N and F represent calculations for $\beta = 0.25$ and 1, respectively, where $\beta \equiv T_e/T_p$ is the ratio of electron to proton temperatures.}
\label{fig:int_ratio}
\end{figure}

We can set theoretical bounds on the ratio of broad to narrow H$\alpha$ emission, $I_b/I_n$, as shown in Figure [\ref{fig:int_ratio}].  Generally, our predictions agree quite well with observations, but a glaring discrepancy persists: the theoretical prediction of $I_b/I_n \sim 0.1$ (i.e., interior-to-surface H$\alpha$ ratio) in SNR 1987A is lower by an order of magnitude compared to the observed ratio.  This points to two possibilities: there is a mechanism for interior emission we have not yet modeled (C. Fransson, Aspen talk, 2007); and/or the assumption of thin shock fronts in these SNRs is a flawed one.

\section{The Shock Transition Zone}

What if these shock fronts that we have been modeling as mathematical discontinuities all along do indeed have a finite width?  We decided to investigate this issue, resolving the atomic physics while keeping the plasma kinetics unresolved \cite{Heng:2007}.  In a system of pre-shock atoms and post-shock ions, there must exist a transition zone in which one population is converted into the other, via charge transfers and ionizations.  This ``shock transition zone'' has a width on the order of the mean free path of atoms passing through the ionized gas, $l_{\rm{zone}} \sim 10^{15} n^{-1}_0$ cm, where $n_0$ is the pre-shock ionic density (in cm$^{-3}$).

\begin{figure}
  \includegraphics[width=5.5in]{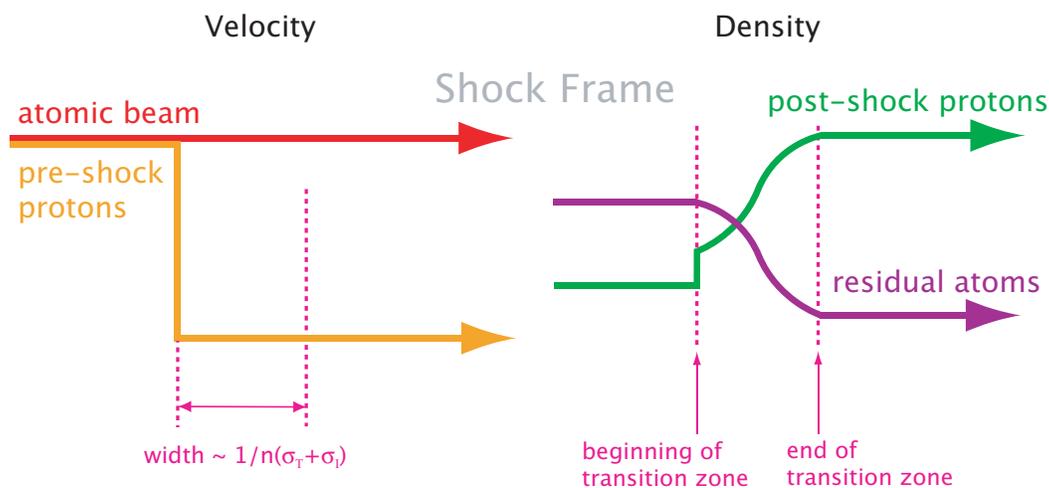}
  \caption{Schematic diagram of the shock transition zone, in the case of a strong shock, taken from \cite{Heng:2007}.  The width of the zone is on the order of the mean free path of interactions (charge transfer and ionization).  The velocity of the ions goes down to 1/4 of its pre-shock value almost immediately, according to the Rankine-Hugoniot jump condition.  The ionic density first jumps by a factor of 4 to conserve momentum, then eventually evolves to a value which depends on the pre-shock ionic density.}
\label{fig:zone}
\end{figure}

The results are surprising --- for a strong ($\sim 1000$ km s$^{-1}$) shock, the ions are shocked immediately.  There is no velocity structure within the shock transition zone (Figure \ref{fig:zone}), thus validating the thin shock assumptions of \cite{CKR80} and \cite{HengMcCray:2007}.  The ionic velocities are decelerated to 1/4 of their pre-shock values at the beginning of the zone, while the ionic densities jump by a factor of 4 to conserve momentum, consistent with the Rankine-Hugoniot jump conditions.  There is, however, structure in both the atomic and ionic densities, which is relevant to the study of Ly$\alpha$ resonant scattering in young SNRs (pre-Sedov-Taylor phase).  The mean free path for the scattering of Ly$\alpha$ photons is much less than $l_{\rm{zone}}$; photons are produced in the zone but scatter in a distance much less than its width.  There is evidence for Ly$\alpha$ resonant scattering in SNR 1987A \cite{Heng:2006}.

\section{The Future}

As SNR 1987A enters its third decade, many questions regarding its fate abound.  A central one concerning the reverse shock is: when will it disappear?  There is a competition between pre-shock atoms crossing the shock --- and ultimately emitting the H$\alpha$ and Ly$\alpha$ photons we observe --- and post-shock (ultraviolet and X-ray) photons diffusing upstream.  These photons are capable of ionizing the atoms before they have a chance to undergo impact excitation (or charge transfer).  If the flux of ionizing photons exceeds that of the atoms, the reverse shock emission will vanish.  In \cite{Smith:2005}, we predict this event to occur between about 2012 and 2014; we are currently planning further observations to finetune this prediction (PI: J. Danziger).  These observations may yet shed light on the origins of the interior emission.  One thing is for certain --- SNR 1987A will provide current and future, young generations of astronomers/astrophysicists (such as myself) with an abundance of rich problems to ponder over.


\begin{theacknowledgments}
I am deeply grateful to Richard McCray for being a wonderful advisor, and to the Aspen Center for Physics for the hospitality of their support staff, and the generosity of both financial support and the {\it Martin and Beate Block Prize} (awarded at the conference).  I thank Roger Chevalier, John Danziger, Eli Dwek, Alak Ray, Dick Manchester, Bryan Gaensler, Brian Metzger, Claes Fransson, Dieter Hartmann, Peter Lundqvist, Karina Kjaer, Alicia Soderberg, Lifan Wang, Philipp Podsiaklowski, Shigeyama Nagataki, Avi Loeb, Bob Kirshner, Saurabh Jha, Jason Pun, Andrew MacFayden, Jeremiah Murphy and Chris Stockdale for intriguing conversations and/or wonderful company during the conference.  I apologize if I have left out anyone who belongs to the preceding list.
\end{theacknowledgments}

\section{Appendix: Deriving $I_b$ and $I_n$ in Heng \& McCray (2007)}
\label{sect:append}

In this section, I derive in more detail the H$\alpha$ broad- and narrow-line rate coefficients, denoted $I_b$(H$\alpha$) and $I_n$(H$\alpha$) respectively, and simply stated in \cite{HengMcCray:2007}.  For simplicity, we refer to them just as $I_b$ and $I_n$.

For narrow H$\alpha$ line emission, atoms are found in a beam and may be excited an arbitrary number of times until it gets transformed into a broad neutral via charge transfer or destroyed by ionization.  Let the probability of excitation be $P_{E_0}$, where the ``0'' means that the atom has undergone zero charge transfers prior to excitation.  Let the rate coefficient for excitation to the atomic level $n$ be $R_{E_0,n}$.  Considering multiple excitations yield:
\begin{equation}
R_{E_0,n} \left( 1 + P_{E_0} + P^2_{E_0} + P^3_{E_0} + ... \right) = R_{E_0,n} ~\sum^\infty_{i=0} P^i_{E_0} = \frac{R_{E_0,n}}{1 - P_{E_0}},
\end{equation}
since $0 < P_{E_0} < 1$.

One can consider excitations up to some level $m$, depending on the atomic data available.  Ignoring collisional de-excitation, the rate coefficient for the narrow H$\alpha$ line is
\begin{equation}
I_n = \frac{C_{32}}{1 - P_{E_0}} ~\sum^m_{n=3} R_{E_0,n} C_{n3},
\end{equation}
where $C_{ij}$ is the probability that an atomic excited to a state $i$ will transit to a state $j<i$ via {\it all} possible cascade routes; it is thus called the ``cascade matrix''.

Let us next derive the rate coefficient for the broad H$\alpha$ line.  We first account for charge transfer to excited states directly from the atomic beam to the level n, which has a rate coefficient $R_{T^*_0,n}$.  Accounting for multiple excitations before such a charge transfer, we have $R_{T^*_0}/(1 - P_{E_0})$.  Next, we need to account for the creation of broad neutrals and the multiple {\it charge transfers} they are capable of undergoing:
\begin{equation}
\frac{P_{T_0}}{1-P_{E_0}} ~\left[1 + \frac{P_T}{1-P_E} +  \left(\frac{P_T}{1-P_E}\right)^2 + ...\right] = \frac{P_{T_0}}{1-P_{E_0}} ~\sum^\infty_{i=0} \left(\frac{P_T}{1-P_E}\right)^i = \frac{P_{T_0}}{P_I} \left(\frac{1 - P_E}{1- P_{E_0}}\right).
\end{equation}
The $1/(1-P_{E_0})$ and $1/(1-P_E)$ terms account for repeated excitations prior to engaging in charge transfer.  As in \cite{HengMcCray:2007}, we make the approximation that the rate coefficients and probabilities are approximately unchanged after the first charge transfer, and thus they do not possess a subscript (e.g., $P_E$ versus $P_{E_0}$).  Physically, these are reactions involving broad neutrals.  Charge transfer to excited states and excitation of the broad neutrals are given by $R_{T*,n}/(1-P_E)$ and $R_{E,n}/(1-P_E)$, respectively.  Putting everything together and summing excitations to some level $m$, we get:
\begin{equation}
I_b = \frac{C_{32}}{1 - P_{E_0}} ~\sum^m_{n=3} \left[ \frac{P_{T_0}}{P_I}\left(R_{E,n} + R_{T^*,n}\right) + R_{T^*_0,n} \right] ~C_{n3}.
\end{equation}


\bibliographystyle{aipproc}   

\bibliography{kh}

\end{document}